\documentclass[12pt]{article}
\usepackage[dvips]{graphicx}
\usepackage[rflt]{floatflt}
\usepackage{mathrsfs}
\usepackage{amssymb}
\usepackage{amsmath}
\usepackage{amsmath}
\usepackage{amssymb}

\begin{document}

\begin{center}\bf
The generalized Kondo lattice model and its spin-polaron
implementation\\ for cuprates by projection method.
\end{center}

\begin{center}
V.\,V.\, Val'kov$^{a)}$,
        D.\,M.\, Dzebisashvili$^{a,b)}$,
        A.\,F.\, Barabanov$^{c)}$
\end{center}

\begin{center}\small\sl
$^{a}$ Institute of Physics, Russian Academy of Sciences, Siberian
Branch, Krasnoyarsk, 660036 Russia\\
$^{b}$ Siberian State Aerospace University, Krasnoyarsk, 660014 Russia\\
$^{c}$ Institute for High Pressure Physics, Moscow oblast,
Troitsk, 142190 Russia
\end{center}

{\small It is shown that the spin-fermion model found to be an
effective low-energy implementation of the three-band Emery model
after Wannier transformation of $p_x$- and $p_y$-orbitals in the
oxygen ions subsystem is reduced to the generalized Kondo lattice
model. Its essential feature is the presence of spin-correlated
hoppings of the current carriers between distant cells. Numerical
calculations of the spin-polaron spectrum demonstrate the
important role of the distant hoppings.}
\bigskip

{\bf 1. Introduction}

The normal phase quasiparticle structure of cuprate
high-temperature superconductors and the nature of the low-energy
interactions are crucial to establish the Cooper instability
mechanism, anomalous temperature dependence of the kinetic
coefficients and a number of other unusual properties
\cite{Keimer2015,Plakida_book_10}. The theory of doped 2D
antiferromagnet being far from a complete is mainly developed
within 2D Hubbard model, generalized t-J-models, three-band Emery
(TBE) model \cite{Emery1987,Varma_1987,Hirsch_1987,EmeryReiter88}
and spin-fermion (SF) model. We will not dwell on the first two
models, but note that they differ significantly from the SF and
TBE models by the fact that their both charge and spin subsystems
are formed by the same carriers. Therefore, we will proceed from
the more realistic, in our view, SF and TBE models, where the spin
system and the system of charge carriers are determined by $d-$
and $p-$ ions respectively.

It is known that in the strong electron correlations regime the
Hamiltonian of the TBE model can be reduced to the spin-fermion
model
\cite{Zaanen-1988,Barabanov1988,Prelovsek_1988,Matsukawa_1989}.
Note that the spin-fermion model even in the simplest
consideration allows one to obtain the basic motive of the
cuprate's hole spectrum \cite{Zaanen-1988,Barabanov1988}, i.e. it
appears that the bottom of the carriers spectrum $E(k)$ is near
the boundary of the magnetic Brillouin zone $E(k)\sim$
$(\cos(k_{x})+\cos(k_{y}))^{2}$.

In this paper the Hamiltonians of the TBE and SF models are
reduced to the generalized Kondo lattice model, which takes into
account the spin-correlated distant hoppings. Based on the
comparison of the spectral dependencies of the spin-polaron
quasiparticles for the generalized Kondo model and its reduced
version (when only the nearest hoppings are left) we argue the
importance of the long-range spin-correlated hoppings.
\bigskip

{\bf 2. The generalized Kondo lattice model}

The Hamiltonian of the Kondo lattice model is derived from TBE
model for the CuO$_2$-plane
\cite{Emery1987,Varma_1987,Hirsch_1987,EmeryReiter88} which in
conventional notation is as follows:
\begin{eqnarray}\label{emery-1}
\hat{\mathscr H}=\hat{\mathscr H}_{pd}+\hat{\mathscr H}_{I},
\end{eqnarray}
where
\begin{eqnarray}\label{Hpd}
\hat{\mathscr H}_{pd}=\sum_{f}\left( \varepsilon
_{d}\hat{n}_{f}^{d}+U_{d}~\hat{n}_{f\uparrow
}^{d}\hat{n}_{f\downarrow }^{d}\right) +\sum_{l}\left( \varepsilon
_{p}\hat{n}_{l}^{p}+U_{p}~\hat{n}_{l\uparrow
}^{p}\hat{n}_{l\downarrow }^{p}\right) +\nonumber\\
+V_{pd}\sum_{f\delta }\hat{n}_{f}^{d}\hat{n}_{f+\delta
}^{p}+t^{pd}\sum_{f\delta \sigma }(\vartheta (\delta)d_{f\sigma
}^{+}~p_{f+\delta ,\sigma }+H.c.)+\hat{T},
\end{eqnarray}
\begin{eqnarray}
\hat{T}=\sum_{l\Delta\sigma}
t\cdot \varrho(\Delta)p_{_{l,\sigma }}^{+}p_{_{l+\Delta,\sigma}},~~~~~~
\hat{\mathscr H}_{I}=\frac{I_{1}}{2}\sum_{fg}S_{f}S_{f+g}
+\frac{I_{2}}{2}\sum_{fd}S_{f}S_{f+d}.
\end{eqnarray}

Here, $\varepsilon _{p}$ and $\varepsilon _{d}$ are energies of a
hole on the oxygen ion O and copper ion Cu respectively, $U_{p}$
and $U_{d}$-- Coulomb energy of two holes on O and Cu ions,
$V_{pd}$-- repulsion energy of two holes at the nearest oxygen and
copper sites, $t_{pd}$ - hybridization parameter between adjacent
oxygen $p-$ and copper $d-$ orbitals. Vectors $l$ correspond to
the positions of O-sites. Vector $\delta$ in (\ref{Hpd}) takes
four values: $\pm\delta_x,~\pm\delta_y$ $(\delta_x=(a/2,0),
\delta_y=(0, a/2))$ and connects nearest O-sites in positions
$f+\delta$ with Cu-site $f$. Function $\vartheta(\delta)$ takes
into account the relationship between the phases of copper and
oxygen orbitals in the hybridization processes. For commonly used
orbitals the function $\vartheta(\delta)$ takes following values:
$\vartheta(\delta)=\mp 1$ at $\delta=(\pm a/2,0), (0,\pm a/2)$.

The term $\hat{T}$ corresponds to the direct hole hoppings between
the nearest oxygen ions with tunneling integral
$t\varrho(\Delta)$. Its sign is defined by the function
$\rho(\Delta)$, which depends on the orientation of the line
connecting oxygen ions between which the hopping occurs.

Vector $\Delta$ runs four values $(\pm a/2,\pm a/2)$ and couple O
ion in site $l$ with the nearest to it O ion with site index
$l+\Delta$. For chosen oxygen orbital's phases $\rho(\Delta)=1$ if
$\Delta=\pm(a/2,a/2)$ and $\rho(\Delta)=-1$ if
$\Delta=\pm(a/2,-a/2)$.

The second term $\hat{\mathscr H}_{I}$ in (\ref{emery-1})
correspond to the superexchange interaction between spins on the
nearest and next-nearest ($d=\pm g_{x}\pm g_{y}$) sites. $\pm
g_{x}=(\pm a,0)$, $\pm g_{y}=(0,\pm a)$ -- vectors connecting
nearest neighbors of the copper lattice. It is convenient to use
frustration parameter $p$ and effective exchange $I$:
$I_{1}=(1-p)I$, $I_{2}=pI$, $0\leq p\leq 1$, $I>0$. In this
approach it is assumed that when considering the properties of the
spin subsystem of copper ions, we can use a two-dimensional (2D)
AFM frustrated Heisenberg model with $S=1/2$. AFM interaction
between nearest copper spins in CuO$_{2}$ plane is large (order of
$0.13$ eV  $\cong 1500K$ and considerably exceeds the interplane
exchange. The interplane exchange is mainly responsible for the
long-range order, which is observed in the insulating phase of
CuO$_{2}$ planes (for La$_{2}$CuO$_{4}$ the characteristic Neel
temperature is $T_{N}\sim 300$ K). However, at relatively low hole
doping the long-range AFM order vanishes in the entire temperature
range. Such behavior is reasonably well simulated by introducing
frustration \cite{idg88}. Cluster calculations indicate the
presence of a sufficiently large frustration parameter
$I_{2}/I_{1}\sim 0.1$ even for the undoped LSCO \cite{amms89}. In
the present work a quantitative analysis of the spin subsystem is
carried out in the framework of a spherically symmetric
self-consistent theory
\cite{KondoYamaji1972,Shimahara1991,BarBer1994}. The subsystem of
spins, localized on copper ions, is considered in a spin-liquid
state which is spherically symmetric in spin space. This means
that the spin correlation functions $C_{r}=\langle
S_{f}S_{f+r}\rangle$ satisfy the conditions $C_{r}=3\langle
S_{f}^{x(y,z)}S_{f+r}^{x(y,z)}\rangle$, $\langle S_{f}^{\alpha
}\rangle =0,(\alpha =x,y,z)$.

In the considered below case $U_{p}=V_{pd}=t=0$, $U_{d}\neq 0$ the
three-band Hamiltonian is considerably simplified
\begin{eqnarray}\label{emery-2}
\hat{\mathscr H}_{pd}=\sum_{f}\left(\sum_{\sigma}\varepsilon
_{d}\hat{n}_{f,\sigma }^{d}+U_{d}~\hat{n}_{f\uparrow
}^{d}\hat{n}_{f\downarrow }^{d}\right)+\sum_{l,\sigma}\varepsilon
_{p}\hat{n}_{l,\sigma }^{p}+t^{pd}\sum_{f,\delta ,\sigma
}(\vartheta (\delta)d_{f,\sigma }^{+}~p_{f+\delta,\sigma }+H.c.).
\end{eqnarray}

This allows for operators $p_{f+\delta_{x},\sigma }^{+}$ and
$p_{f+\delta_y,\sigma}^+$, written in $k$-representation, by
performing a unitary transformation to go to the new operators
$\psi_{k,\sigma}^+$ and $\phi_{k,\sigma}^+$ such that
$\psi_{f,\sigma}^+$ does not hybridise to the Cu-orbitals.

Let us introduce operators
\cite{Shastry_1989,Jefferson_1992,Gavrichkov_1998,Digor_2002}:
\begin{eqnarray}\label{unit-k}
& \phi _{k\sigma }=\left( s_{kx}~p_{k,\delta _{x},\sigma
}+s_{ky}~p_{k,\delta _{y},\sigma }\right) /s_{k},\ \psi _{k\sigma
}=\left( s_{ky}~p_{k,\delta _{x},\sigma }-s_{kx}~p_{k,\delta
_{y},\sigma }\right)
/s_{k},\nonumber\\
& p_{k,\delta _{x},\sigma
}=\frac{1}{\sqrt{N}}\sum_{f}e^{-ikf}p_{f+\delta_x,\sigma
},~~p_{k,\delta _{y},\sigma
}=\frac{1}{\sqrt{N}}\sum_{f}e^{-ikf}p_{f+\delta
_{y},\sigma},\nonumber\\
& s_{kx}=\sin \frac{k_{x}}{2},~~s_{ky}=\sin
\frac{k_{y}}{2},~~s_{k}=\sqrt{s_{kx}^{2}+s_{ky}^{2}}.
\end{eqnarray}

Hamiltonian (\ref{emery-2}) in the representation of operators
$\psi_{f,\sigma}^+$, $\phi_{f,\sigma}^+$ reads:
\begin{eqnarray}
\hat{\mathscr H}_{pd}=\varepsilon _{d}\sum_{f,\sigma
}\hat{n}_{f,\sigma }^{d}+U_{d}\sum_{f}\hat{n}_{f\uparrow
}^{d}\hat{n}_{f\downarrow }^{d}+\varepsilon _{p}\sum_{f,\sigma
}(\phi _{f,\sigma }^{+}\phi _{f,\sigma }+\psi _{f,\sigma }^{+}\psi
_{f,\sigma }),\nonumber\\
\hat{V}=-2it^{pd}\Sigma_{f,g,\sigma }(s_{f-g}~d_{f\sigma }^{+}\phi
_{g\sigma }-s_{f-g}^{\ast }~\phi _{g\sigma }^{+}d_{f,\sigma }),
\end{eqnarray}
where $s_{f-g}=\frac{1}{N}\sum_{k}e^{ik(f-g)}s_{k}$.

In the second-order perturbation theory in $t^{pd}$ the
Hamiltonian $\hat{\mathscr H}_{pd}$ is represented as:
\begin{eqnarray}\label{Hfi_eff1}
\hat{\mathscr H}_{eff}^{\phi }=N(\varepsilon _{d}-4\tau
)+\widetilde{\varepsilon }_{\phi }\sum_{f,\sigma }\phi _{f,\sigma
}^{+}\phi _{f,\sigma }+\varepsilon _{p}\sum_{f,\sigma }\psi
_{f,\sigma }^{+}\psi _{f,\sigma }-{\large t_{g}}\sum_{f,g,\sigma
}\phi _{f,\sigma }^{+}\phi _{f+g,\sigma }+
\nonumber\\
+4\tau (1+\eta )\sum_{f,n,m,\sigma }\left( s_{n}s_{m}^{\ast
}\right) \phi _{f-m,\sigma }^{+}\Sigma _{\mathbf{\sigma
}_{1},\alpha }S_{f}^{\alpha }\hat{\sigma}_{\mathbf{\sigma \sigma
}_{1}}^{\alpha }\phi _{f-n,\sigma_{1}}~,
\end{eqnarray}
where notations are introduced:
\begin{eqnarray}
\tau
=\frac{(t^{pd})^{2}}{\varepsilon_{pd}},~~
\varepsilon_{pd}=\varepsilon_{p}-\varepsilon_{d},~~\eta
=\frac{\varepsilon_{pd}}{U_{d}-\varepsilon_{pd}},~~\widetilde{\varepsilon
}_{\phi }= \varepsilon _{p}+2\tau (1-\eta),~~t_{g}=\frac{\tau
}{2}(1-\eta ).
\end{eqnarray}
At appropriate for cuprates values of the model parameters
\cite{Ogata2008}: $t^{pd}=1.3$ eV, $\varepsilon_{pd}=3.6$ eV and
$U_d=10.5$ eV we get: $\tau=0.47$ eV, $\eta=0.52$, $t_g=0.11$ eV.

The overlapping parameters $(s_{n}s_{m}^*)$ are determined only by
the geometry of the lattice and decrease rapidly with the increase
of the number of coordination sphere in the copper ions lattice
${\large n,m}$:
\begin{eqnarray}\label{ss-overlaps}
(s_{0}s_{0}^*)=0.920,~(s_{1}s_{0}^*)=(s_{g}s_{0}^*)=-0.136,~
(s_{2}s_{0}^*)=(s_{d}s_{0}^*)=-0.022,
\nonumber\\
(s_{3}s_{0}^*)=(s_{2g}s_{0}^*)=-0.010,~~(s_{1}s_{1}^*)=
(s_{g}s_{g}^*)=0.020,~(s_{2}s_{1}^*)=0.003.
\end{eqnarray}

The last term in (\ref{Hfi_eff1}) describes the current carriers
scattering on localized spins followed by spin flipping, i.e. the
Kondo scattering processes. Besides they involve only
$\phi$-fermion states.

We show that these processes, considering rapid decay of the
parameters (\ref{ss-overlaps}), define the low-energy part of the
excitation spectrum, enable to exclude from consideration the
$\psi$-carriers and lead to the justification of the spin-polaron
concept. We omit the constant in equation (\ref{Hfi_eff1}) and
express it as a sum of single-site terms and terms describing the
$\phi$-holes motion:
\begin{eqnarray}\label{Hfi_eff2}
\hat{\mathscr H}_{eff}^{\phi }=\widetilde{\varepsilon}_{\phi
}\sum_{f,\sigma}\phi_{f,\sigma}^{+}\phi_{f,\sigma
}+\widehat{J}_{\phi-d}+\varepsilon _{p}\sum_{f,\sigma}\psi
_{f,\sigma}^+\psi_{f,\sigma}+\hat{t}_g^{\phi
}+\hat{t}^{SC},\nonumber\\
\widehat{J}_{\phi -d}=J_{\phi-d}\sum_{f,\sigma }\phi_{f,\sigma
}^{+}\Sigma_{\sigma_{1},\alpha}S_{f}^\alpha\hat{\sigma}_{\sigma\sigma_{1}}^{\alpha
}\phi _{f,\sigma_1}=2J_{\phi -d}\sum_{f}\vec S_f{\vec
s_{f}}^{~\phi },
\nonumber\\
J_{\phi-d}=4\tau (1+\eta)(s_{0}s_{0}^*),~~\hat
t_{g}^{\phi}=-{\large t_{g}}\sum_{f,g,\sigma
}\phi_{f,\sigma}^+\phi_{f+g,\sigma },
\nonumber\\
\hat{t}^{SC}=4\tau(1+\eta)\sum_{f,n,m,\sigma }(1-\delta
_{m,0}\delta _{n,0})\left( s_{n}s_{m}^*\right) \phi
_{f-m,\sigma}^{+}\Sigma _{\sigma_1,\alpha }S_{f}^{\alpha
}\hat{\sigma}_{\sigma\sigma_1}^{\alpha }\phi_{f-n,\sigma_{1}}.
\end{eqnarray}

In the expression for $\hat{t}^{SC}$ the  factor
$(1-\delta_{m,0}\delta_{n,0})$ ensures that under the sum there is
no the term in which all three sites are the same, i.e.
$\hat{t}^{SC}$ corresponds to the motion of a hole with spin
flipping.

The biggest in the effective Hamiltonian (\ref{Hfi_eff2}) is the
$\phi-d$-exchange on-site interaction $\widehat{J}_{\phi-d}$ with
parameter $2J_{\phi-d}=8\tau (1+\eta)(s_{0}s_{0}^*)$. It leads to
the formation of two energy levels at each site. The lower level
corresponds to $\phi-d$ singlet state, with energy
$\widetilde{\varepsilon}_\varphi^-=\widetilde{\varepsilon
}_{\phi}-\frac32J_{\phi-d}$ and with wave function
$|\varphi_{f}\rangle=\varphi_f^+|O\rangle$, where
$\varphi_f^+=\frac{1}{\sqrt{2}}(\phi_{f+}^+
Z_f^{-0}+Z_f^{+0}\phi_{f,-}^+)$. For the localized spins we use
Hubbard operators $Z_f^{\lambda_1\lambda_2}$, where $|O\rangle$
stands for a vacuum in one-site cluster. The upper three-fold
degenerate level $\widetilde{\varepsilon }_{\chi
}^{+}=\widetilde{\varepsilon }_\phi+\frac{1}{2}J_{\phi-d}$
corresponds to the three $\phi-d$ triplet states with wave
functions:
\begin{eqnarray}
|\chi _{f,m}\rangle =\chi_{f,m}^{+}|O\rangle,~~~m=-1,0,1,
\end{eqnarray}
$$\chi_{f,0}^+=\frac{1}{\sqrt{2}}(\phi_{f,+}^+Z_f^{-0}-Z_f^{+0}\phi_{f,-}^+),~
\chi_{f,+1}^{+}=\phi_{f,+}^+Z_f^{+0},~
\chi_{f,-1}^+=\phi_{f,-1}^+Z_f^{-0}.$$ The discussed level
splitting is qualitatively similar to that which occurs in
considering the Zhang-Rice polaron \cite{Zhang_Rice_1988}.

To justify the spin polaron concept we first define the separation
 $\Delta _{\varepsilon _{p}=\varepsilon _{\psi
};\widetilde{\varepsilon }_{\varphi }^{-}}=\varepsilon
_{p}-\widetilde{\varepsilon }_{\varphi }^-$ between $\varepsilon
_{\psi }$ and $\widetilde{\varepsilon }_{\varphi }^{-}$ levels:
\begin{eqnarray}
\Delta _{\varepsilon _{p}=\varepsilon _{\psi
};\widetilde{\varepsilon }_{\varphi }^{-}}=\varepsilon
_{p}-\widetilde{\varepsilon }_{\varphi }^{-}=\varepsilon
_{p}-\widetilde{\varepsilon }_{\phi }+\frac{3}{2}J_{\phi
-d}\approx 2\tau [ 5+7\eta ].
\end{eqnarray}

The low-frequency part of the hole spectrum will be formed near
the lower one-site level $\widetilde{\varepsilon}_{\varphi}^-$ and
should be determined primarily by the motion of the polaron state
$|\varphi_{f}\rangle \rightarrow |\varphi_{f+g}\rangle$. If the
half-width of the band $W$ of this movement due to terms $\hat
t_g^{\phi}$ and $\hat{t}^{SC}$ (\ref{Hfi_eff2}) will be less than
$\Delta_{\varepsilon_{p}=\varepsilon _{\psi
};\widetilde{\varepsilon}_{\varphi}^-}$, then we are allowed to
omit from consideration the $\psi$ carriers.

The bandwidth $2W$ of this motion should indicate a polaron
narrowing and, in any case, must be proportional to the spin-spin
correlation function $C_{g}=\langle \vec S_{f}\vec S_{f+g}\rangle
\simeq 0.2\div 0.3$ (a typical value for 2D AFM in the spin-liquid
state). This is evident from the fact that the motion
$|\varphi_f\rangle\rightarrow |\varphi_{f+g}\rangle$ is always
associated with a combination of operators
$\widehat{\varphi}_{f+g}^+\widehat{\varphi}_f$, containing the
spin operators at the neighboring sites.

The half-width of the band $W_g$, generated by the term
$\hat{t}_g^\phi$ (\ref{Hfi_eff2}), is estimated as:
$$W_{g}=2{\large t_{g}Ñ_{g}=2}\frac{\tau}{2}(1-\eta)Ñ_{g}=
\tau (1-\eta)Ñ_{g}\ll \Delta_{\varepsilon _{p}=\varepsilon _{\psi
};\widetilde{\varepsilon}_{\varphi}^{-}}\approx 2\tau \lbrack
5+7\eta ],$$ i.e. $W_{g}\ll \Delta _{\varepsilon_{p}=\varepsilon
_{\psi};\widetilde{\varepsilon}_{\varphi }^-}$.

The half-width of the band $W_{SC}$, induced by $\hat{t}^{SC}$ in
(\ref{Hfi_eff2}), is defined by the term:
\begin{eqnarray}
\hat{t}_{g}^{SC}=t_g^{SC}[\sum_{f,g,\sigma}\phi_{f+g,\sigma
}^{+}\Sigma _{\sigma_1,\alpha}S_{f}^{\alpha}\hat{\sigma}_{\sigma
\sigma_1}^{\alpha }\phi_{f,\sigma_1}+\sum_{f,g,\sigma }\phi
_{f,\sigma }^{+}\Sigma _{\sigma_1,\alpha}S_{f}^{\alpha
}\hat{\sigma}_{\sigma \sigma_1}^{\alpha }\phi
_{f+g,\sigma_1}]= \nonumber\\
=t_g^{SC}[\sum_{f,g,\sigma}\phi_{f+g,\sigma }^{+}\Sigma
_{\mathbf{\sigma}_1,\alpha}S_{f}^{\alpha
}\hat{\sigma}_{\sigma\sigma_1}^{\alpha
}\phi_{f,\sigma_1}+H.c.],~~~t_{g}^{SC}=4\tau (1+\eta
)(s_{0}s_{g}^*),
\end{eqnarray}
which in $\hat{t}^{SC}$ contains the maximum overlapping
$(s_{0}s_{g}^*)=-0.136$ (see. formula (\ref{ss-overlaps})).
Parameter $t_{g}^{SC}$ 13 times smaller than the constant of
$\phi-d$-exchange coupling $2J_{\phi-d}$. Taking into account the
factor $C_g$ this leads to inequality
\begin{eqnarray}
W_{SC}\ll \Delta _{\varepsilon _{p}=\varepsilon _{\psi
};\widetilde{\varepsilon }_{\varphi }^{-}}.
\end{eqnarray}

Thus we have: $W_{g}\ll \Delta _{\varepsilon _{p}=\varepsilon
_{\psi };\widetilde{\varepsilon }_{\varphi }^{-}}$, $W_{SC}\ll
\Delta _{\varepsilon _{p}=\varepsilon _{\psi
};\widetilde{\varepsilon }_{\varphi }^{-}}$, which are the
consequence of inequalities $J_{\phi -d}\gg t^{SC}\gg {\large
t_{g}}$. These conditions mean that in the model (\ref{Hfi_eff2}),
we can omit the term $\varepsilon_{p}\sum_{f,\sigma}\psi
_{f,\sigma}^+\psi_{f,\sigma}$ and introduce the spin-polaron
concept.

In the approximation of a small number of $\phi$-holes after
performing transformation to the Hubbard projection operators
$\phi_{f,\sigma }^{+}\Rightarrow X_{f}^{\sigma 0}$, the
Hamiltonian (\ref{Hfi_eff2}) takes the form of the Kondo lattice
Hamiltonian:
\begin{eqnarray}\label{HK_eff}
\hat{\mathscr H}_{eff}^{K}=\ \widetilde{\varepsilon }_{\phi
}\sum_{f\sigma}X_{f}^{\sigma 0}X_{f}^{0\sigma }+ 2J_{\phi
-d}\sum_{f}\vec S_f{\vec s_f}^{~\phi}-t_g\sum_{f,g,\sigma
}X_{f}^{\sigma
0}X_{f+g}^{0\sigma }\ + \nonumber\\
+4\tau (1+\eta )\sum_{f,m,n,\sigma }(1-\delta _{m,0}\delta
_{n,0})\left( s_{n}s_{m}^{\ast }\right) X_{f-m}^{\sigma 0}(\Sigma
_{\sigma_1,\alpha }S_{f}^{\alpha }\hat{\sigma}_{\sigma
\sigma_1}^{\alpha }X_{f-n}^{0\sigma _{1}}).
\end{eqnarray}

If in the expression (\ref{HK_eff}) the distant spin-correlated
hoppings in the first approximation are ignored (i.e. only the
terms with ($s_{0}s_{g}^*$) are left), we arrive at the
Hamiltonian of the reduced Kondo lattice model:
\begin{eqnarray}\label{HKr_eff}
\hat{\mathscr
H}_{eff}^{Kr}=\widetilde{\varepsilon}_\phi\sum_{f,\sigma
}X_{f}^{\sigma 0}X_{f}^{0\sigma }+ 2J_{\phi-d}\sum_{f}{\vec
S}_{f}{\vec s_{f}}^{~\phi}-{\large t_{g}}\sum_{f,g,\sigma
}X_{f}^{\sigma
0}X_{f+g}^{0\sigma }+\nonumber\\
+t^{SC}_g[\sum_{f,g,\sigma}\phi_{f+g,\sigma}^+\Sigma_{\sigma_1,\alpha}S_f^{\alpha
}\hat{\sigma}_{\sigma\sigma_1}^\alpha\phi_{f,\sigma_1}+H.c.].
\end{eqnarray}
On the basis of this model in the works
\cite{Ramsak_1989,Ramsak_1990} the spectral properties of the
Fermi quasiparticles in cuprate superconductors were studied.

The description of excitations corresponding to the Hamiltonian
(\ref{HK_eff}) is convenient to carry out in the framework of
Zwanzig-Mori projection method \cite{Zwanzig1961,Mori1965}.
\bigskip

{\bf 3. Basis operators and projection method}

Let's introduce three sets of basis operators of charge
excitations $A_{j,f,\sigma}$ $(j=1,2,3)$, and their Fourier
transforms $A_{k,f,\sigma}$:
\begin{eqnarray}\label{Basis_1}
& A_{1,f,\sigma }=X_{f}^{0\sigma },~~A_{2,f,\sigma }=\Sigma
_{\sigma_1,\alpha }S_{f}^{\alpha
}\hat{\sigma}_{\sigma\sigma_1}^{\alpha }X_{f}^{0\sigma
_{1}},~~~A_{3,f,\sigma}=\Sigma_{\sigma_1,\alpha }S_{f}^{\alpha
}\hat{\sigma}_{\sigma\sigma_1}^{\alpha}
X_{f+g}^{0\sigma_1},\nonumber\\
& A_{j,k,\sigma }=N^{-1/2}\sum_{f}e^{-ikf}A_{j,f,\sigma }.
\end{eqnarray}
Next we consider, related to these operators, the two-time
retarded Green's function $G_{ij}(k,t)$ and its Fourier transforms
$G_{ij}(k,\omega )=\langle \langle A_{ki}|A_{kj}^{+}\rangle
\rangle _{\omega }$ ($i,j=1,2,3$).

To close the equations of motion for $\langle\langle
A_{ki}|A_{kj}^+\rangle\rangle_\omega$:
\begin{eqnarray}
\omega \langle \langle A_{ki}|A_{kj}^{+}\rangle \rangle _{\omega
}=K_{ij}+\langle \langle \lbrack A_{ki},\hat{\mathscr
H}_{eff}^{Kr}~]|A_{kj}^{+}\rangle \rangle _{\omega },
\end{eqnarray}
it is required to calculate the energy matrix: $D_{ij}=\langle
\{[A_{ik},\hat{\mathscr H}_{eff}^{Kr}],A_{jk}^{+}\}\rangle$, and
the matrix: $K_{ij}=\langle\{A_{ik},A_{jk}^{+}\}\rangle$. Then the
Fermi Green's functions are obtained from the set of equations
having matrix form $G=\left(\omega -DK^{-1}\right)^{-1}K$, and the
spectrum of Fermi excitations is determined by the poles of the
Green's functions: $G_{ij}(k,\omega
)=\sum_{n=1}^{3}\frac{z_{(i,j)}^{n}(k)}{\omega -E_{n}(k)}$,
$(i,j=1,2,3)$.

Thus, the problem of finding spectrum $E_{n}(k)$ and residues
$z_{(i,j)}^{n}(k)$ is reduced to the calculation of the matrix
elements $K_{ij}$ and $D_{ij}$. To calculate these elements we
denote the terms in the Hamiltonian (\ref{HKr_eff}), contributing
to the matrix $D$, by: $\hat{J}$, $\hat{t}$, $\hat{\tau}^{(SC)}$
and $\hat{\varepsilon}$.

Using relation
\begin{equation}
\langle \tilde{S}_{f_{1}}\tilde{S}_{f_{2}}\tilde{S}_{f_{3}}\rangle
=-\delta _{f_{1},f_{2}}C_{f_{1}-f_{3}}-\delta
_{f_{2},f_{3}}C_{f_{1}-f_{2}}+\delta
_{f_{1},f_{3}}C_{f_{1}-f_{2}},
\end{equation}
that is fulfilled when averaged over a singlet state of the
undoped CuO$_{2}$-plane, as well as the identity
\begin{equation} \sum_{f}s_{f-n}s_{f-m}^{\ast }=\delta
_{m,n}-\frac{1}{4}\sum_{g}\delta (m-n-g),
\end{equation}
where $g$ runs over nearest neighbors, we find the expressions for
the matrix elements of operators of the matrix $D$
($D_{ij}=J_{ij}+t_{ij}+\tau^{(SC)}_{ij}+\varepsilon_{ij}=D_{ji}$):
\begin{equation}\label{Dij}
J_{11}=0;~J_{12}=J\frac{3}{8};~~J_{31}=2JC_{1}\gamma
_{1};~~J_{22}=-J\frac{3}{8};~~J_{32}=-2JC_{1}\gamma
_{1};~~J_{33}=2JC_{1};
\end{equation}$$
~\tau _{11}^{(SC)}=0;~~\tau _{12}^{(SC)}=t^{SC}(3+4C1)\gamma
_{1};~~\tau _{13}^{(SC)}=4~t^{SC}\left( \frac{3}{4}+4C_{1}\gamma
_{1}^{2}+2C_{2}\gamma _{2}+C_{3}\gamma _{3}\right) ;
$$$$
\tau _{22}^{(SC)}=-t^{SC}~8~C_{1}\gamma _{1};~~~
\tau_{23}^{(SC)}=-4~t^{SC}\left( \frac{3}{4}-C_{1}+2C_{2}\gamma
_{2}+C_{3}\gamma _{3}\right) ;~~~\tau
_{33}^{(SC)}=-t^{SC}~32~C_{1}\gamma _{1};
$$$$
t_{11}=-4~t~\gamma _{1};~~t_{22}=-4~t~C_{1}\gamma
_{1};~~t_{32}=-t~(3+8C_{2}\gamma _{2}+4C_{3}\gamma _{3});
$$$$
t_{33}=-4~t~(9C_{1}\gamma_{1}
+6C_{4}\gamma_{4}+C_{6}\gamma_{6});~~t_{12}=t_{13}=0;
$$$$
\varepsilon _{11}=\bar{\varepsilon}_{p};~~~
\varepsilon_{22}=\bar{\varepsilon}_{p}\frac{3}{4};~~~
\varepsilon_{32}=\bar{\varepsilon}_{p}~4~C_{1}~\gamma_{1};~~~
\varepsilon_{12}=
\varepsilon_{13}=0,~~\varepsilon_{33}=4\bar{\varepsilon}_{p}\left(
\frac{3}{4}+2C_{2}\gamma_{2}+C_{3}\gamma_{3}\right),
$$
where $~~\gamma _{1}=(\cos (k_{x})+\cos (k_{y}))/2, ~~ \gamma
_{2}=\cos (k_{x})\cos (k_{y}), ~~ \gamma _{3}=(\cos (2k_{x})+\cos
(2k_{y}))/2, $ $$ \gamma _{4}=(\cos (2k_{x})\cos (k_{y})+\cos
(2k_{y})\cos (k_{x}))/2, ~~ \gamma _{6}=(\cos (3k_{x})+\cos
(3k_{y}))/2.
$$
For matrix elements of the matrix $K$ we have:
\begin{equation}
K_{11}=1,~~K_{12}=K_{13}=0,~~K_{22}=\frac{3}{4},~~
K_{32}=4C_1\gamma_{1},~~K_{33}=3+8~C_{2}\gamma_{2}+4C_{3}\gamma_{3}.
\end{equation}

A similar procedure may be used for unreduced Hamiltonian
(\ref{HK_eff}). Let us represent matrix elements of the
corresponding energy matrix in the form: $D^K_{ij}={\mathscr
J}_{ij}+t_{ij}+\varepsilon_{ij}$, where terms $t_{ij}$ and
$\varepsilon_{ij}$ equal to the ones calculated earlier in
(\ref{Dij}), and element ${\mathscr J}_{ij}$ takes into account
all the interactions in $\hat{\mathscr H}^K_{eff}$ due to
localized spin operators $S^{x(y,z)}_f$. The calculation of
${\mathscr J}_{ij}$ ($i,j=1,2,3$) yields ($\lambda_f=\sum_g
s_{f+g}$):
\begin{eqnarray}
{\mathscr J}_{11}=0,~~{\mathscr J}_{21}=4\tau(1+\eta)s_k \sum_f
e^{ikf} s^*_f C_f, ~~{\mathscr J}_{31}=4\tau(1+\eta)s_k \sum_f
e^{ikf} \lambda^*_f C_f,
\nonumber\\
{\mathscr J}_{22}=4\tau(1+\eta)\sum_f\left(-e^{-ikf}s_f s^*_0
-e^{ikf}s_0 s^*_f+|s_f|^2 \right)C_f,
\nonumber\\
{\mathscr J}_{32}=4\tau(1+\eta)\sum_f\left(-e^{-ikf}s_f\lambda^*_0
-e^{ikf}s_0\lambda^*_f+s_f\lambda^*_f \right)C_f,
\nonumber\\
{\mathscr
J}_{33}=4\tau(1+\eta)\sum_f\left(-e^{-ikf}\lambda_f\lambda^*_0
-e^{ikf}\lambda_0\lambda^*_f+|\lambda_f|^2 \right)C_f.
\end{eqnarray}
Note that, as before, all the matrix elements $D^K_{ij}$ in the
low-density limit can be expressed only in terms of the spin-spin
correlation functions.

Let us return to the question of how to choose the operator basis
(\ref{Basis_1}). Generally speaking, the intuitive natural choice
of the basis is dictated by the equations of motion for the
$\langle \langle A_{ki}|A_{kj}^+\rangle\rangle_{\omega}$, when on
the first step the "bare"\ hole operator $X_{f}^{0\sigma }$ is
chosen as a basis operator $A_{ki}$. As a result of commutation
$[X_{f}^{0\sigma },\hat{\mathscr H}_{eff}^{Kr}]$ there appear new
types of operators like $\Sigma_{\mathbf{\sigma}_1,\alpha
}S_{f}^{\alpha }\hat{\sigma}_{\sigma\sigma_1}^\alpha
X_{f+g}^{0\sigma_1}$, $\Sigma_{\sigma_1,\alpha}S_f^\alpha
\hat{\sigma}_{\sigma\sigma_1}^{\alpha}X_{f}^{0\sigma_1}$. These
operators are included in the basis and for them the equations of
motion are written again. This procedure continues until a certain
stage. In the last step there are always operators beyond the
basis. These operators are projected on to the already selected
basis.

To clarify the question of what kind of excitation operators
remain out of the basis, consider the case of $\hat{\mathscr
H}=\hat{J}=J\sum_{\sigma,\sigma_1}X^{\sigma 0}(\vec
S\frac{1}{2}\vec \sigma_{\sigma\sigma_1})X^{0\sigma_1}$. Here and
below the site index is omitted. A complete set of operators
describing excitations from the two states $|\uparrow\rangle$,
$|\downarrow\rangle$ without holes to the four possible states
with a hole (a singlet and three triplet states) yields eight
transitions operators. These operators can be classified according
to the irreducible representations of the rotation group. As a
result, there appear two doublets $\widehat{s1}_{\sigma}^+$,
$\widehat{s2}_{\sigma}^+$ and a quartet
$\widehat{q}_{-3/2}^{+},\dots,\widehat{q}_{3/2}^{+}$, where
$\widehat{s1}_{\sigma}^+$ describes a transition to the singlet
with a hole, and $\widehat{s2}_{\sigma}^+$ -- to the triplet
state. The quartet is always related to the triplet state
transitions. Operators of these excitations are of the form
($\gamma=\pm 1=2\sigma$):
\begin{eqnarray}
& \widehat{s1}_{\sigma }^{+}=\frac{1}{\sqrt{2}}(X^{\gamma
0}Z^{\overline{\gamma}\overline{\gamma }}-X^{\overline{\gamma
}0}Z^{\gamma\overline{\gamma}}),~~ \widehat{s2}_{\sigma
}^{+}=\frac{1}{\sqrt{2}}(2X^{\gamma 0}-\ X^{\gamma
0}Z^{\overline{\gamma }\overline{\gamma }}+X^{\overline{\gamma
}0}Z^{\gamma \overline{\gamma
}}),\nonumber\\
& \widehat{q}_{\frac{3}{2}}^{~+}=X^{+0}Z^{++},~
\widehat{q}_{\frac{1}{2}}^{~+}=X^{-0}Z^{+-}
+X^{+0}Z^{--}-X^{+0}Z^{++}.
\end{eqnarray}

Here we made a transformation to the Hubbard operators for the
localized spins. Linear combinations of $\widehat{s1}_{\sigma
}^{+}$\ and $\widehat{s2}_{\sigma }^{+}$ yield $A_{1,\sigma
}=X^{0\sigma }$ and $A_{2,\sigma }=\Sigma _{\mathbf{\sigma
}_{1},\alpha }S^{\alpha }\hat{\sigma}_{\mathbf{\sigma \sigma
}_{1}}^{\alpha }X^{0\sigma _{1}}$, besides:
\begin{eqnarray}
[\widehat{s1}_\sigma,\hat{J}]=-\frac{3}{4}J\cdot\widehat{s1}_\sigma,~
[\widehat{s2}_\sigma,\hat{J}]=\frac{J}{4}\cdot\widehat{s2}_{\sigma},~
[\widehat{q}_{\frac{3}{2}},\hat{J}]=\frac{J}{4}\cdot\widehat{q}_{\frac{3}{2}},~
[\widehat{q}_{\frac{1}{2}},\hat{J}]=\frac{J}{4}\cdot\widehat{q}_{\frac{1}{2}}.
\end{eqnarray}
Thus beyond our basis are four quartet operators
$\widehat{q}_{-3/2}^{+},\dots,\widehat{q}_{3/2}^{+}$.

A similar group analysis can be carried out for the two-site
excitation operators and can be used to determine selection rules
for the matrix elements.
\bigskip

{\bf 4. Results and discussion}

\begin{figure}[h]
\begin{center}
\includegraphics[width=380pt, height=320pt, angle=0, keepaspectratio]{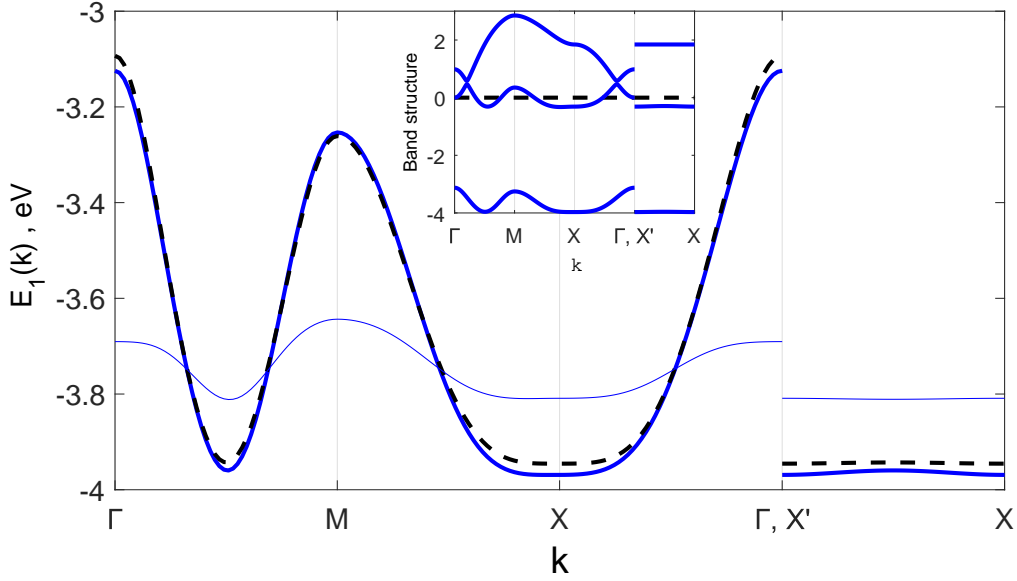}
\caption{\label{Fig1} The dispersion curves of the spin-polaron
excitations in the lower band for the three effective Hamiltonians
of the Emery model. The solid bold curve is for the generalized
Kondo lattice Hamiltonian (\ref{HK_eff}), a thin solid line stands
for the reduced Kondo lattice model (\ref{HKr_eff}) and the dashed
curve is the energy spectrum in the spin-fermion model. The
calculations were performed for the following values of the pair
spin-spin correlation functions: $C_1=-0.255$, $C_2=0.075$,
$C_3=0.064$ and all $C_j$ with $j>3$ are zero. For simplicity, we
also put $I=0$. Symmetry points of the Brillouin zone are:
$\Gamma=(0,0)$, $M=(\pi,\pi)$, $X=(\pi,0)$, $X'=(0,\pi)$. The
inset shows the dispersion curves describing three solutions of
the cubic dispersion equation $\omega-DK^{-1}=0$ for the effective
Hamiltonian (\ref{HK_eff}) in the basis (\ref{Basis_1}). The lower
curve coincides with the bold solid curve of the main figure.
Horizontal dashed line in the inset shows the inactive
$\psi$-orbital energy level.}
\end{center}
\end{figure}

Figure \ref{Fig1} shows the lower band dispersion curves of the
spin-polaron excitations $E_1(k)$, obtained by solving the
dispersion equation $\omega-DK^{-1}=0$ for the three effective
Hamiltonians of the Emery model. The solid bold line is calculated
for the full effective Hamiltonian (\ref{HK_eff}) using three
operators basis (\ref{Basis_1}). The solid thin line represents
the excitation spectrum for the reduced effective Hamiltonian
(\ref{HKr_eff}) with the same basis operators. It is evident that
in the $\Gamma-M$-direction both curves show a dispersion minimum
nearby the point ($\pi/2,\pi/2$) of the Brillouin zone. The
essential difference between these curves is the Fermi excitations
bandwidth. Analysis of the solid curves in figure \ref{Fig1} shows
that neglect of the long-range spin-correlated hole hoppings leads
to a significant (almost five times) reduction in the spin-polaron
bandwidth.

For comparison, the dashed curve shows the spin-polarons
dispersion calculated for the effective spin-polaron Hamiltonian
(\ref{Hfi_eff1}), written in terms of $p$-orbitals (see e.g.
\cite{Barabanov_JETP2001, DVB_2013}). In this case the used three
basis operators are:
\begin{eqnarray}\label{Basis_2}
A_{1,f,\sigma}=p_{f+\delta_x,\sigma},~~A_{2,f,\sigma}=
p_{f+\delta_y,\sigma},~~A_{3,f,\sigma}=\Sigma
_{\delta,\sigma_1,\alpha }S_f^{\alpha
}\hat{\sigma}_{\sigma\sigma_1}^\alpha p_{f+\delta,\sigma}.
\end{eqnarray}
It can be seen that the spin-polaron spectrum obtained earlier in
the spin-fermion model is well reproduced by the Kondo lattice
model (\ref{HK_eff}) and is not reproduced by the reduced Kondo
lattice model (\ref{HKr_eff}).

In the inset in the figure \ref{Fig1} by solid lines are shown the
three dispersion curves related to the unreduced Kondo-lattice
model (\ref{HK_eff}) calculated in the basis of the three
operators (\ref {Basis_1}). The horizontal dashed line corresponds
to the inactive $\psi$ orbital energy. It can be seen that the
lower spin-polaron band is separated from the bare energy of the
oxygen $p$-orbital $\varepsilon_p$ (accepted here as zero) down by
about 3 eV.

An important feature of the hole spectrum in the cuprate
high-temperature superconductors is the absolute minimum nearby
the $(\pi/2,\pi/2)$-point of the Brillouin zone. The dispersion
curves in Figure \ref{Fig1} exhibit the minimum only in the
direction $\Gamma-M$, but not in the direction $M-M'$. This
"incorrect"\ behavior is due to the fact that when deriving the
effective Hamiltonian (\ref{HK_eff}) direct p-p hoppings were
rejected. Accounting for these hoppings obviously results in
renormalization of the tunneling integral between nearest
neighbors $t_g$, as well as the induction of the new hoppings
between distant cells, which intensity is rapidly decreases with
increasing the distance.

Consider for example the, largest of the induced, hoppings between
next-nearest cells with the tunneling integral $t_d$. Related to
these processes kinetic energy operator
\begin{eqnarray}
\hat t^d= -t_d\sum_{f,d,\sigma}\phi_{f\sigma}^+\phi_{f+d,\sigma}
\end{eqnarray}
should be added to the Hamiltonians (\ref{HK_eff}) and
(\ref{HKr_eff}). Thereafter each matrix element $D_{ij}$ from
(\ref{Dij}) is renormalized additively on the value of $t^d_{ij}$
$(i,j=1,2,3$). Calculation of the matrix elements
$t^d_{ij}~(=t^d_{ji})$ gives:
\begin{eqnarray}
& t^d_{11}= -4t^d\gamma_2,~~t^d_{12}=t^d_{13}=0,~~t^d_{22}= -4t^d
C_2\gamma_2,~~t^d_{23}= -8t^d\left(C_1\gamma_1+C_4\gamma_4\right),\nonumber\\
& t^d_{33}=-8t^d\left(3/4 + 3C_2\gamma_2 + 2C_3\gamma_3 +
C_5\gamma_5 +C_7\gamma_7 \right),
\end{eqnarray}
where $ \gamma_5=\cos 2k_x\cos 2k_y$, $\gamma_7=(\cos k_x\cos 3k_y
+ \cos 3k_x\cos k_y)/2$.

\begin{figure}[h]
\begin{center}
\includegraphics[width=420pt, height=290pt, angle=0, keepaspectratio]{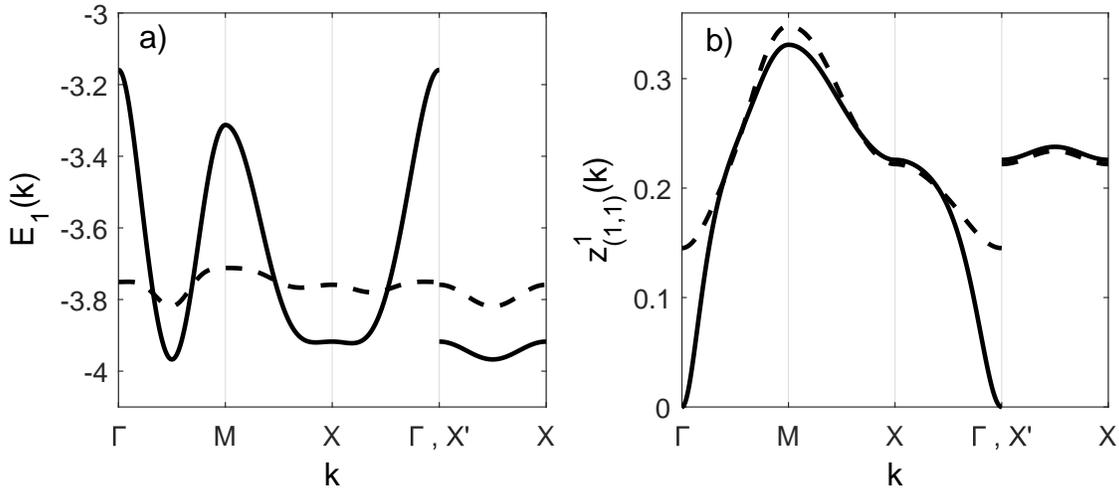}
\caption{\label{Fig2} The spectrum (a) and spectral intensity (b)
of the lower spin-polaron band calculated taking into account
hoppings between next-nearest neighbors. The solid lines
correspond to the full generalized Kondo lattice model
(\ref{HK_eff}), the dashed line --- to the reduced Kondo lattice
model (\ref{HKr_eff}). In both cases the three operators basis
(\ref{Basis_1}) has been used. Tunneling integral $t_d=0.05$ eV.
The rest parameters of the model are the same as in Figure
\ref{Fig1}.}
\end{center}
\end{figure}

In the figure \ref{Fig2}a by solid line is shown the spin polarons
spectrum calculated in the three operators basis (\ref{Basis_1})
for the full (unreduced) Kondo lattice Hamiltonian (\ref{HK_eff})
taking into account hoppings between next-nearest neighbors. The
dashed line in the same figure shows the spin-polaron spectrum in
the reduced Kondo lattice model (\ref{HKr_eff}), obtained in the
same basis (\ref{Basis_1}) and taking into account the tunneling
operator $\hat t^d$.
Comparison with the similar curves in figure \ref{Fig1} shows that
the inclusion of p-p-hoppings leads to a minimum in the Fermi
excitations spectrum near the ($\pm\pi/2,\pm\pi/2$)- points of the
Brillouin zone, both in the direction $\Gamma-M$ and in the
direction $X-X'$. This important fact gives rise at these points
to the small hole pockets, which are observed in the experiments
on angle-resolved photoelectron spectroscopy in the lightly hole
doped high-temperature cuprate superconductors.

Figure \ref{Fig2}b shows the wave vector dependence of the
residues $z^{(1)}_{(1,1)}(k)$ of the Green's function
$G_{11}(k,\omega)$. These residues determine the contribution of
the "bare" holes states to the lower spin-polaron state for each
value of $k$. It is seen that in the case of the full Kondo
lattice Hamiltonian (\ref{HK_eff}) the "bare"\ holes contribution
to the spin-polaron state vanishes at $\Gamma$-point of the
Brillouin zone. This behavior is completely consistent with the
calculations of the function $z^{(1)}_{(1,1)}(k)$ within the
spin-fermion model (i.e., within the effective Hamiltonian
(\ref{Hfi_eff1}) written in the representation of the initial
oxygen $p$ orbitals) using operator basis (\ref{Basis_2})
\cite{Kuzian_PRB1998,VDB_2015}. The mentioned feature, however, is
not reproduced for the reduced Hamiltonian (\ref{HKr_eff}). In
this case, as follows from the figure \ref{Fig2}b, in the $\Gamma$
-point of the Brillouin zone the Green's function residue
$z^{(1)}_{(1,1)}(k)$ does not vanish. This fact once again
demonstrates the importance of taking into account the distant
spin-correlated hoppings.
\bigskip

{\bf 5. Conclusion}

We obtained an effective low-energy Hamiltonian of the three-band
Emery model in the form of a generalized Kondo lattice model. An
important feature of this Hamiltonian is that it retains the
spin-correlated hoppings between distant cells.

Within the generalized Kondo lattice model we analyzed the role of
the long-range spin-correlated hoppings which are usually
discarded in particular calculations. Comparing the dispersion
curves of the spin-polaron spectrum and the spectral density of
the "bare"\ holes calculated for the generalized Kondo lattice
model (\ref{HK_eff}) and the reduced Kondo lattice model
(\ref{HKr_eff}) the essential role of the long-range
spin-correlated hoppings is demonstrated. In particular, we show
that the retaining of these interactions leads to a significant
increase in the spin-polaron bandwidth as well as to an additional
reduction of the spin-polaron's minimal energy.

The role of direct oxygen-oxygen hoppings was also analyzed. It
was shown that these hoppings are necessary to take into account
to reproduce the experimentally observed minimum in the spectrum
of spin-polaron excitations in the ($\pm\pi/2,\pm\pi/2$)-points of
the Brillouin zone.

Note that earlier in the Kondo lattice model for two-dimensional
doped antiferromagnets the pseudogap behavior of the current
carrier's spectral function and anomalous temperature dependence
of the kinetic coefficients were considered
\cite{Barabanov_2010,Larionov_2014}. However, these studies
contained a significant drawback due to ignorance of the hole
motion processes with spin-flipping. As a result, in order to
achieve a satisfactory agreement between theory and experiment it
was necessary to artificially introduce additional hoppings on the
first three nearest neighbors, so that the bare band bottom (i.e.,
without interaction with the spins) was close to the magnetic
Brillouin zone boundary. This problem doesn't arise if the low
temperature properties of cuprates are studied within the obtained
in the work generalized Kondo lattice model (\ref{HK_eff}).
\bigskip

This work was supported by RFBR (grants 16-02-00073 and
16-02-00304), as well as a Complex program of Siberian Branch of
Russian Academy of Sciences II.2P (grant 0358-2015-0005).


\begin{thebibliography}{99}

\bibitem{Keimer2015} B. Keimer, S. A.
Kivelson, M. R. Norman, S. Uchida, J. Zaanen, NATURE, v.{\bf 518},
p.179 (2015).

\bibitem{Plakida_book_10}
Nikolay Plakida, High-Temperature Cuprate Supercoductors
Experiment, Theory, and Applications, Springer, 570 pp., 2010.

\bibitem{Emery1987} V. J. Emery, Phys.Rev.Lett. \textbf{58}, 2794 (1987).

\bibitem{Varma_1987} C.M. Varma, S. Schmitt-Rink, E. Abrahams, Solid State
Commun. \textbf{62} 681 (1987).

\bibitem{Hirsch_1987} J.E. Hirsch, Phys.Rev.Lett. \textbf{59}, 228
(1987).

\bibitem{EmeryReiter88} V.J. Emery, G. Reiter, Phys.Rev.B {\bf 38},
4547 (1988); {\bf 38}, 11938 (1988).

\bibitem{Zaanen-1988} J. Zaanen, A.M. Oles, Phys.Rev.B {\bf 37}, 9423
(1988).

\bibitem{Barabanov1988} A.F. Barabanov, L.A. Maximov, G.V.
Uimin, JETP Lett., {\bf 47}, 622 (1988).

\bibitem{Prelovsek_1988} P. Prelovsek, Physics Letters A.
{\bf 126}, 287 (1988).

\bibitem{Matsukawa_1989} Í. Matsukawa, H. Fukuyama ,J.Phys.Soc.Jpn. {\bf 58},
2845 (1989).

\bibitem{idg88} M. Inui, S. Doniach and M. Gabay, Phys.Rev.B {\bf 38},
6631 (1988).

\bibitem{amms89} J.F. Annet, R.M. Martin, A.K. McMahan, S. Satpathy,
Phys.Rev.B {\bf 40}, 2620 (1989).

\bibitem{KondoYamaji1972} J. Kondo, K. Yamaji, Prog. Theor. Phys.,
\textbf{47}, 807 (1972).

\bibitem{Shimahara1991} H. Shimahara, S. Takada, J. Phys. Soc. Jpn.,
\textbf{60}, 2394 (1991).

\bibitem{BarBer1994} A.F. Barabanov, V.M. Berezovskii,
JETP, {\bf 79}, 627 (1994).

\bibitem{Shastry_1989} B.S. Shastry,
Phys. Rev. Lett. {\bf 63}, 1288 (1989).

\bibitem{Jefferson_1992} J.H. Jefferson, H. Eskes, L.F. Feiner,
Phys. Rev. B {\bf 45}, 7959 (1992).

\bibitem{Gavrichkov_1998} V.A. Gavrichkov, S.G. Ovchinnikov,
Physics of the Solid State, {\bf 40}, 184 (1998).

\bibitem{Digor_2002} D.F. Digor, V.A. Moskalenko, Theor.Math.Phys., {\bf 130}, 320 (2002).

\bibitem{Ogata2008} M. Ogata, H. Fukuyama, Rep.Prog.Phys. \textbf{71},
036501 (2008).

\bibitem{Zhang_Rice_1988} F.C. Zhang, T.M. Rice, Phys.Rev.B \textbf{37},
3759 (1988).

\bibitem{Ramsak_1989} A. Ramsak, P. Prelovsek, Phys.Rev.B {\bf 40}, 2239 (1989).

\bibitem{Ramsak_1990} A. Ramsak, P. Prelovsek, Phys.Rev.B {\bf 42}, 10415 (1990).

\bibitem{Zwanzig1961} R. Zwanzig, Phys. Rev. \textbf{124}, 983 (1961).

\bibitem{Mori1965} H.Mori, Prog. Theor. Phys. \textbf{33}, 423 (1965).

\bibitem{Barabanov_JETP2001}
A.F. Barabanov, A.A. Kovalev, O.V. Urazaev, A.M. Belemuk, JETP
{\bf 92}, 677 (2001); [Rus. ZhETF {\bf 119}, 777 (2001)].

\bibitem{DVB_2013} D.M. Dzebisashvili, V.V. Val'kov, A.F. Barabanov,
JETP Lett. {\bf 98}, 596 (2013).

\bibitem{Kuzian_PRB1998} R.O. Kuzian, R.Hayn,  A.F. Barabanov, L.A. Maksimov,
Pys. Rev. B {\bf 58}, 6194 (1998).

\bibitem{VDB_2015} V.V. Val'kov, D.M. Dzebisashvili, A.F. Barabanov, Phys.
Lett. A. \textbf{379}, 421 (2015).

\bibitem{Barabanov_2010} A.F. Barabanov, A.M. Belemouk,
JETP, {\bf 111}, 258 (2010).

\bibitem{Larionov_2014} I.A. Larionov, A.F. Barabanov, JETP Lett., {\bf
100}, 811 (2014).

\end{thebibliography}
\end{document}